\documentclass[a4paper,11pt]{article}
\usepackage{aaskaiid}
\usepackage{orcidlink}

\title{The Impact of Dense RM Grids on the Study of Intra-cluster and Intra-group Magnetic Fields}
\ShortTitle{The impact of dense RM grids}

\author[1]{Francesca Loi\orcidlink{0000-0002-8627-6627}}
\ShortName{Loi et al.} 
\author[1]{Valentina Vacca\orcidlink{0000-0003-1997-0771}}
\author[2]{Shane P.~O'Sullivan\orcidlink{0000-0002-3968-3051}}
\author[3]{Craig Anderson\orcidlink{0000-0002-6243-7879}}
\author[4]{Chiara Stuardi\orcidlink{0000-0003-1619-3479}}
\author[1]{Federica Govoni\orcidlink{0000-0003-3644-3084}}
\author[1]{Matteo Murgia\orcidlink{0000-0002-4800-0806}}
\author[5]{Annalisa Bonafede\orcidlink{0000-0002-5068-4581}}
\author[3]{Ettore Carretti\orcidlink{0000-0002-3973-8403}}
\author[1]{Filippo M. Maccagni\orcidlink{0000-0002-9930-1844}}
\author[6,7]{Tessa Versnstrom\orcidlink{0000-0001-7093-3875}}

\affiliation[1]{INAF - Osservatorio Astronomico di Cagliari,              via della scienza 5, Selargius, Italy}
\emailAdd{francesca.loi@inaf.it}
\affiliation[2]{Departamento de Física de la Tierra y Astrofísica \& IPARCOS-UCM, Universidad Complutense de Madrid, 28040 Madrid, Spain}
\affiliation[3]{Research School of Astronomy \& Astrophysics, The Australian National University, Canberra ACT 2611, Australia}
\affiliation[4]{INAF - Istituto di Radioastronomia (IRA), Via Gobetti 101, 40129 Bologna, Italy}

\affiliation[5]{DIFA - Università di Bologna, via Gobetti 93/2, I-40129 Bologna, Italy}

\affiliation[6]{Australia Telescope National Facility, CSIRO, Space and Astronomy, PO Box 1130, Bentley WA 6102,
Australia}

\affiliation[7]{International Centre for Radio Astronomy Research, University of Western Australia, 7 Fairway, Crawley, WA 6009, Australia}

\abstract{The presence of diffuse radio sources in galaxy clusters and the recent discovery of polarized signals associated with the tails of a jellyfish galaxy indicates that intra-cluster/intra-group magnetic fields can influence the physics of these environments and the evolution of the embedded galaxies. A better reconstruction of the properties of such fields is therefore fundamental to understand in detail the physical processes in galaxy groups and clusters and the evolution of the embedded sources. The SKAO represents a great opportunity to perform these studies through the analysis of the so-called rotation measure (RM) grid, since polarization properties of radio sources are modified by the intervening magnetic field.
In this manuscript, we illustrate the prediction on the density of the RM grid considering the SKA-mid polarization survey planned by the SKA Magnetism Science Working Group. Moreover, we describe how it is possible to measure intra-cluster/intra-group magnetic fields with the RM grid. Eventually, we quantify the improvement in the precision and accuracy of the magnetic field measurements compared to what is achievable with current surveys such as the POSSUM survey.}

\begin{document}
\newcommand{\actaa}{Acta Astron.} 
\newcommand{\araa}{ARA\&A} 
\newcommand{\aar}{A\&ARv} 
\newcommand{\aapr}{A\&ARv} 
\newcommand{\ab}{Astrobiol.} 
\newcommand{\aj}{AJ} 
\newcommand{\apj}{ApJ} 
\newcommand{\apjl}{ApJL} 
\newcommand{\apjs}{ApJSS} 
\newcommand{\ao}{Appl. Opt.} 
\newcommand{\apss}{Astro. \& Space Sci.} 
\newcommand{\aap}{A\&A} 
\newcommand{\aaps}{A\&AS.} 
\newcommand{\baas}{Bull. Am. Astron. Soc.} 
\newcommand{\caa}{Chinese A\&A} 
\newcommand{\cjaa}{Chinese J. A\&A} 
\newcommand{\cqg}{Class. Quantum Gravity} 
\newcommand{\gal}{Galaxies} 
\newcommand{\gca}{Geo. Cosmo. Acta} 
\newcommand{\icarus}{Icarus} 
\newcommand{\jcap}{JCAP} 
\newcommand{\jgr}{J. Geophys. Res.} 
\newcommand{\jgrp}{J. Geophys. Res. Planets} 
\newcommand{\jqsrt}{J. Quant. Spectrosc. Radiat. Transf.} 
\newcommand{\memsai}{Mem. SAIt} 
\newcommand{\mnras}{MNRAS} 
\newcommand{\nat}{Nature} 
\newcommand{\nastro}{Nat. Astron.} 
\newcommand{\ncomms}{Nat. Commun.} 
\newcommand{\nphys}{Nat. Phys.} 
\newcommand{\na}{New Astron.} 
\newcommand{\nar}{New Astron. Rev.} 
\newcommand{\physrep}{Phys. Rep.} 
\newcommand{\pra}{Phys. Rev. A} 
\newcommand{\prb}{Phys. Rev. B} 
\newcommand{\prc}{Phys. Rev. C} 
\newcommand{\prd}{Phys. Rev. D} 
\newcommand{\pre}{Phys. Rev. E} 
\newcommand{\prx}{Phys. Rev. X} 
\newcommand{\prl}{Phys. Rev. Let.} 
\newcommand{\psj}{Planet. Sci. J.} 
\newcommand{\planss}{Planet. Space Sci.} 
\newcommand{\pnas}{Proc. Natl Acad. Sci. USA} 
\newcommand{\procspie}{Proc. SPIE} 
\newcommand{\pasa}{PASA} 
\newcommand{\pasj}{PASJ} 
\newcommand{\pasp}{PASP} 
\newcommand{\rmxaa}{RMXAA} 
\newcommand{\sci}{Science} 
\newcommand{\sciadv}{Sci. Adv.} 
\newcommand{\solphys}{Sol. Phys.} 
\newcommand{\sovast}{Soviet Ast.} 
\newcommand{\ssr}{Space Sci. Rev.} 
\newcommand{\uni}{Universe} 

\setlength{\bibsep}{0.0pt} 
\maketitle

\section{Introduction}
Galaxy clusters and groups represent the largest gravitationally bound systems in the Universe and are key laboratories for studying the interplay between baryonic matter, relativistic particles, and magnetic fields. The presence of diffuse, extended, synchrotron radio sources such as halos, relics, and mini-halos provides compelling evidence for the existence of large-scale magnetic fields permeating the intracluster medium (ICM) \citep[see ][for a recent review]{vanweeren2019}. 

The measurement of magnetic field properties in cosmic environments, particularly within galaxy clusters and groups, relies heavily on the analysis of rotation measure (RM) \citep[e.g.][]{govoni2004}. This technique utilizes the Faraday effect, where the plane of linear polarization of radio emission is rotated as it passes through a magnetized plasma. Assuming that the rotation is external to the emitting medium, the RM is determined by observing the variation of the polarization angle with the square of the wavelength $\lambda$, captured by the relation: 
\begin{equation}
    \chi(\lambda)=\chi_0 + RM \cdot\lambda^2,
	\label{eq:fd}
\end{equation}
where $\chi(\lambda)$ is the observed polarization angle and $\chi_0 $ the intrinsic angle. The RM is defined as:
\begin{equation}
    RM \propto \int_0^L B_{||} n_e dl,
\label{eq:rm}
\end{equation}
with $n_e$ being the thermal plasma density, $B_{||}$ the line-of-sight parallel component of the intervening magnetic field, and L the distance between the observer and the source. 
By constructing RM maps from high-resolution, wide-band (or multi-frequency) radio interferometric data, it is possible to infer key properties of magnetic fields, such as their strength, orientation, and degree of turbulence, within the intracluster medium. Indeed, detailed RM studies of specific targets, including radio galaxies, cluster relics, and halos, provide crucial constraints on both the local and global characteristics of magnetic fields. Typical RM values associated with intra-cluster and intra-group media range between tens and hundreds rad/m$^2$. Therefore such studies are best performed at GHz frequency, where the Faraday effect induces significant rotations (exceeding 50 degrees) while keeping intra-channel effects negligible (less than a few degrees for MHz-resolution data).
Indeed, GHz frequencies are a compromise between $\sim$100 MHz and $\geq$10-20\,GHz observations. The former are characterized by strong depolarization effects in the case of dense and highly magnetized environments, such as galaxy clusters, and are best suited to reveal the RM associated with filaments of the cosmic web \citep[see][]{Vacca02.2026.SKA, OSullivan01.2026.SKA}. The latter are not susceptible to the Faraday effect caused by the magneto-ionic medius associated with clusters, yet they can be important to observe the intrinsic polarized properties of bright radio sources.\\
Numerical methods are crucial to infer magnetic field properties \citep[see e.g.][]{murgia2004}: starting from a set of magnetic field models and assuming a thermal plasma density distribution derived from X-ray observations, numerical tools compute the RM integral and compare the results to RM images, with the goal of identifying the best-fit model.\\ 
In recent years, this approach has yielded a detailed characterization of magnetic field properties in several systems \citep{murgia2004, govoni2006, guidetti2008, laing2008, bonafede2010, guidetti2010, vacca2010A&A...514A..71V, vacca2012, govoni2017, stuardi2021, Vacca2022MNRAS.514.4969V, Pagliotta2025A&A...700A.139P}, enabling the detection of turbulent magnetic fields fluctuating on scales from a few kiloparsecs to several hundred kiloparsecs. The magnetic field power spectrum is often modeled as a power law:
\begin{equation}
    |B_k|^2 \propto k^{-n},
\end{equation}
with the wavenumber $k=2\pi/\Lambda$, where $\Lambda$ is the physical scale, and the slope $n$ typically ranges between 2 and 4 \citep{murgia2004, govoni2006, guidetti2008, bonafede2010, vacca2010A&A...514A..71V, vacca2012, govoni2017}. In some cases, the power spectrum is better modeled as a broken power law, with a different value of $n$ above a characteristic break wavenumber $k_{break}$ \citep{laing2008, guidetti2010, Vacca2022MNRAS.514.4969V}. In both scenarios, it is assumed that the magnetic field energy is negligible below a minimum $\Lambda_{min}$ and above a maximum scale $\Lambda_{max}$. In recent works \citep{Stuardi2019,Pagliotta2025A&A...700A.139P}, the power spectrum of the magnetic field was imposed from the results of a cosmological magneto-hydro-dynamical (MHD) simulation \citep{paola2019MNRAS.486..623D} that show curved spectra peaking at hundreds of kpc with a k$^{3/2}$ trend at large spatial scales and a decreasing power going at smaller scales, with a different slope according to the cluster dynamical state. \\
Magnetic field strengths typically range from a few to several tens of $\mu$G at the centers of merging and relaxed galaxy clusters, respectively, and decrease towards the periphery, following:
\begin{equation}
    B(r) = B_0 \left ( \frac{n_e(r)}{n_0} \right )^{\eta},
\label{eq:b}
\end{equation}
where $B(r)$ and $n_e(r)$ are the magnetic field strength and thermal plasma density at a distance $r$ from the cluster center, and $B_0$ and $n_0$ are their central values. So far, in targeted studies the parameter $\eta$ has been found to range between 0.5 and $\sim$1, whereas statistical studies showed values of $\eta$ < 0.5 \citep{Osinga2025A&A...694A..44O}.\\
To date, these findings are based on a relatively small sample of galaxy clusters, typically analyzed using RM values from typically one and at most ten cluster-embedded and/or background sources. This limitation is due in part to observational challenges that hinder the detection of polarized signals from faint sources, including beam depolarization (when the magnetic field and the thermal plasma density are tangled on scales smaller than the observing beam), as well as frequency-channel depolarization caused by the Faraday effect. In addition, internal depolarization within magneto-ionic media, such as radio halos, limits polarized emission detection over large scales, which is crucial for probing magnetic fields across a wide range of scales. Nonetheless, numerical simulations suggest that, in selected cases, polarized emission from radio halos can be detected \citep{Govoni2013A&A...554A.102G, loirm2019MNRAS.490.4841L, Vacca2024A&A...691A.334V}, and indeed recent observations have begun to overcome some depolarization-related challenges \citep{Vacca2022MNRAS.514.4969V}. \\
In this context, the role of the SKA precursor and pathfinder is remarkable. \cite{Anderson2021} carried out the first detailed RM study of the Fornax cluster with the ASKAP telescope during the Early Science phase of the Polarization Sky Survey of the Universe’s Magnetism \citep[POSSUM,][]{Gaensler2025PASA...42...91G}. With a 10 hours observation they covered an area of $\sim$34 deg$^2$ between 747 and 1027 MHz, reaching a 30 $\mu$Jy/beam of sensitivity with a resolution of 10$\times$14\, arcsec$^2$, and detected an average density of 25 polarized sources per deg$^2$. A later study of the same target was carried out within the MeerKAT Fornax Survey \citep[MFS,][]{Serra2023A&A...673A.146S}, that consists in 91 pointing of 10 hours each. The density of polarized sources detected between 900 and 1400 MHz with a resolution of 13 arcsec and a sensitivity of 3 $\mu$Jy/beam consists in 80 polarized sources per deg$^2$ \citep{Loi2025A&A...694A.125L}. Such studies demonstrated that a dense RM grid is essential to understand the cluster physics and to reveal phenomena such as matter accretion from the large scale structure.\\
However, deep studies on single targets require long exposure times. A step forward is therefore represented by stacking experiments where RM stacking technique are used to probe magnetized gas in galaxy clusters \citep{Osinga2025A&A...694A..44O} and in and beyond galaxy groups \citep{Anderson2024}.\\
Even if these works represent a huge step forward in the study of large scale magnetic fields, they cannot allow us to obtain a detailed reconstruction, which would require spatially-resolved studies up to high redshift to account for the complex physics at play in this environment and to constrain the magnetization process.\\
In this regards, the advent of the Square Kilometre Array Observatory (SKAO) marks a major leap forward thanks to the expected sensitivity and resolution. The SKA-mid polarization survey, suggested by the SKA Magnetism Science Working Group \citep{Heald2020Galax...8...53H}, could deliver an unprecedentedly dense RM grid, enabling magnetic field studies at levels of precision unattainable with current facilities. \\
The aim of this work is to provide updated values for the RM grid density (Section 2) considering the anticipated performance of SKA-mid AA4  \citep{braun2019anticipatedperformancesquarekilometre} and to develop a framework for assessing the telescope’s potential. We consider two observational strategies for the SKA-mid: a mosaic of pointings with 15-minute integration and a targeted observation with a 10-hour exposure centered on a galaxy cluster. These setups are compared with two recent dense RM grid studies conducted with ASKAP and MeerKAT. Assuming an intra-cluster magnetic field and a thermal plasma model, we generate RM images that account for the sensitivity and resolution of all four observational configurations (Section 3). We then derive the observables traditionally used to constrain magnetic field properties, namely the RM structure function and RM radial profiles (Section 4). Finally, we summarized and discuss the results (Section 5).\\
Throughout this paper we assume a $\Lambda$CDM cosmology with H$_0$ = 71 km\,s$^{-1}$Mpc$^{-1}$, $\Omega_m$=0.3, and $\Omega_{\Lambda}$= 0.7. At z=0.0283, the redshift of the simulated target, 1\,arcsec corresponds to 0.56 kpc.

\section{Current surveys and predictions for RM Grid Density}
In this Section we show the predictions for the RM grid density considering a mosaic and a targeted observations performed with the SKA-mid telescope using the band2 receiver that can acquire spectro-polarimetric signals between 0.95 and 1.76 GHz. We investigate a mosaic of 15 minutes observations per pointing, following the requirement reported in \citep{Heald2020Galax...8...53H}, and a 10 hours targeted observation.   
We determined the observational main parameters using the SKAO sensitivity calculator for mid frequency\footnote{https://sensitivity-calculator.skao.int/mid} considering the AA4 design, assuming that only 30\% of the bandwidth will contribute to the sensitivity, and assuming that such value is representative of the expected noise in Q and U Stokes parameters. In the case of the mosaic, we multiply the sensitivity by a factor of 0.72 to include the contribution of nearby pointings, assuming for the forthcoming analysis that we are investigating a field of view (FoV) with a uniform noise. In both the cases we assumed the Briggs weighting scheme with robust = 0. We obtained a sensitivity $\sigma_{QU}$ of 2.2 and 0.5 $\mu$Jy/beam in the case of the SKA-mid mosaic and targeted observation, respectively, above the expected confusion limit in polarization \citep{loiconfusion2019MNRAS.485.5285L}. The beam size is 0.84$\times$0.73 arcsec$^2$. All the parameters are summarized in Table~\ref{tab:skamid}.\\
\begin{table}[h]
	\centering
	\caption{SKA-mid observations. To compute the sensitivity we consider the contribution of 30\% of the bandwidth and a factor of 0.72 in the case of the mosaic.}
	\label{tab:skamid}
	\begin{tabular}{lccccr} 
		\hline
		Survey & Bandwidth & robust & obs. time & $\sigma_{QU}$ & beam \\
		       & GHz & & [per point.] &[$\muup$Jy/beam] & [arcsec$^2$]  \\
		\hline
		SKA-mid mosaic & 0.95--1.76 & 0 & 15 min & 2.2 & 0.84$\times$0.73\\
		SKA-mid targeted & 0.95--1.76 & 0 & 10 hrs & 0.5 & 0.84$\times$0.73\\
		\hline
	\end{tabular}
\end{table}

We can compute the expected number N of polarized sources per square degree using the equation reported in \cite{loiconfusion2019MNRAS.485.5285L}:
\begin{equation}
 \rm   N(>p/deg^2) = 2.0 \cdot\left ( \frac{p_{1.4\,GHz}}{mJy} \right )^{-0.89},
\end{equation}
where $p_{1.4\,GHz}$ is the limit polarized intensity. This Equation can correctly reproduce the number of polarized source density detected in the MFS, that covers a frequency range similar to the SKA-mid band2, and is within a factor of two of other survey results \citep{rudnick2014,Vanderwoude2024AJ....167..226V,Gaensler2025PASA...42...91G}. 
We consider as the lowest detectable polarized intensity signal $p_{1.4\,GHz}$=5$\sigma_{QU}$.\\
To better investigate the role of the SKA-mid in this science topic we decided to include the densest RM grid reconstructions available at similar frequencies. These have been obtained within POSSUM by \cite{Vanderwoude2024AJ....167..226V} with the ASKAP telescope, and within the MFS. All the relevant numbers are reported in Table~\ref{tab:rmgrid}. In the case of the SKA-mid mosaic, we obtained a denser RM grid with respect to the 60--90 polarized source per square degree reported by \cite{Heald2020Galax...8...53H}. This is due to the improved sensitivity expected for the SKA-mid AA4 design. \\
\begin{table}[h]
	\centering
	\caption{Density of polarized sources. The noise associated with the POSSUM observation is not reported in \citep{Vanderwoude2024AJ....167..226V}.}
	\label{tab:rmgrid}
	\begin{tabular}{lcccr} 
		\hline
		Survey & Bandwidth & $\sigma_{QU}$  & source density& resolution\\
		       & GHz & [$\muup$Jy/beam] & [number/deg$^2$] & [arcsec]  \\
		\hline
		POSSUM & 0.8--1.439 & - & 37 & 21 \\
		MFS & 0.9--1.4 & 3 & 80 & 13 \\
		SKA-mid mosaic & 0.95--1.76 & 2.2 & 110& $\simeq$1  \\
		SKA-mid targeted & 0.95--1.76 & 0.5 & 414 & $\simeq$1  \\
		\hline
	\end{tabular}
\end{table}

\section{Simulated RM images}
\begin{figure}[h]
    \centering
	\includegraphics[width=1\columnwidth]{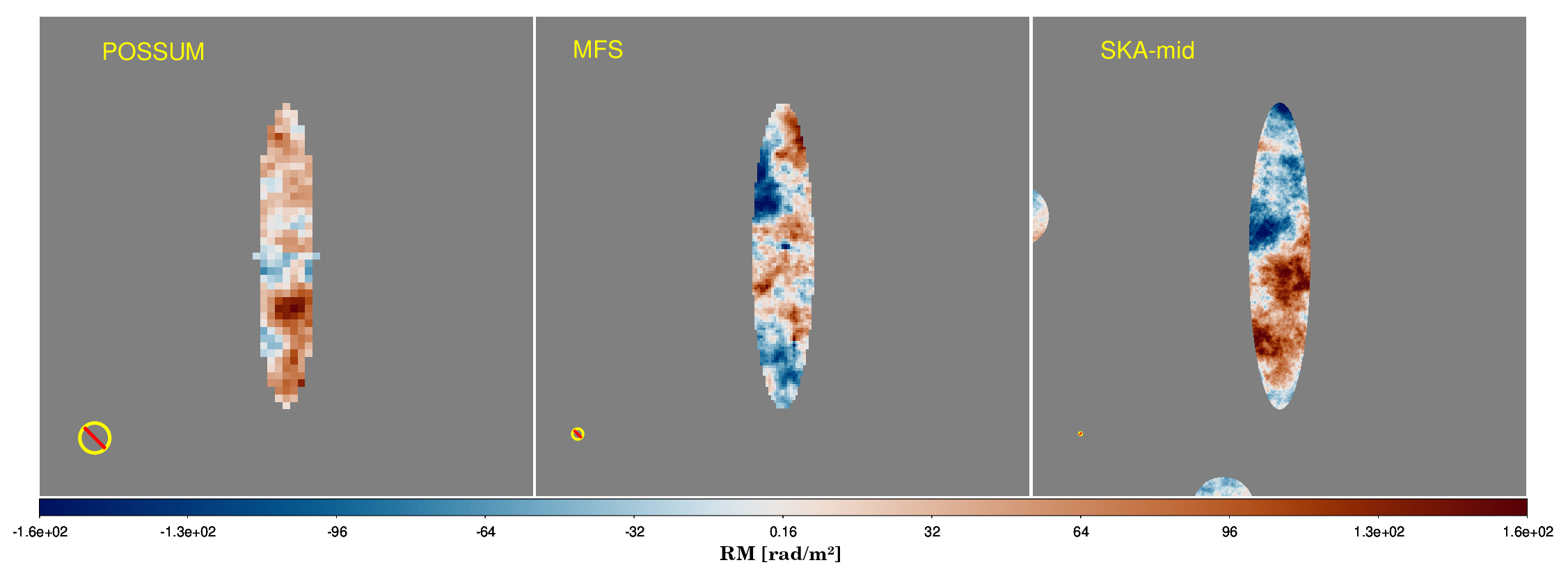}
    \caption{Zoom-in of the simulated RM images centered on the central extended radio source. The beam resolution is shown in the bottom left corner of each panel.}
    \label{fig:simrm}
\end{figure}

Taking into account the resolution and the polarized source density of the four observations listed in Table~\ref{tab:rmgrid}, we made four simulated RM images using the FARADAY tool \citep{murgia2004}. We assumed a magnetic field power spectrum with slope $n$=3, fluctuating between 0.5 and 100\,kpc, and a $\beta$-model for the thermal plasma density:
\begin{equation}
    n_e(r) = n_0 \left (1+\frac{r^2}{r_c^2} \right )^{-3\beta/2},
\end{equation}
with core radius $r_c$=200\,kpc, central density $n_0$= 10$^{-3}$cm$^{-3}$, and radial decrease $\beta$=0.8. We assumed $B_0$=1$\mu$G and $\eta$=1 (see Eq. \ref{eq:b}).\\
We derived the RM from the modeled intra-cluster magnetic field and thermal plasma following the procedure reported in \cite{govoni2017}. We produce three set of RM simulations using a different FoV and pixel size in order to match the resolutions associated with the three telescopes reported in Table \ref{tab:rmgrid}.
With the goal to cover a FoV of at least 1\,deg$^2$ corresponding to 2$\times$2 Mpc$^2$ at the distance of the simulated cluster, and to sample the beam resolution with 4 pixel, we set the cell size to 5 arcsec/pixel, 3 arcsec/pixel and 0.25 arcsec/pixel. In all the images 1 arcsec corresponds to 0.56\,kpc, therefore 1 pixel corresponds to 2.8 kpc, 1.68 kpc and 0.14 kpc in the simulated POSSUM, MFS, and SKA-mid images, respectively.\\
To account for the different numbers of polarized sources expected across the simulated observations, we randomly generate a set of N (x,y) coordinates, where N equals the product of the polarized source density reported in Table \ref{tab:rmgrid} and the FoV of the observation. To enhance the realism of the simulation and adequately sample the smallest scales, we include an elliptical central source with dimensions of 20\,kpc$\times$100\,kpc, two resolved circular sources with a radius of 10\,kpc located within half of the FoV, and the remaining sources divided equally between unresolved sources with beam-sizes and resolved circular sources with a radius of 10\,kpc. We blanked the images by cross-referencing the simulated source catalog; pixels falling outside the defined boundaries of the $N$ generated sources were set to NaN.\\
Fig. \ref{fig:simrm} displays the central region of the resulting RM images, clearly illustrating the resolution improvement anticipated for SKA-mid. The lower resolution of the MFS and the POSSUM simulated images smooth the RM fluctuations, limiting the reconstruction of the magnetic field power spectrum.

\section{Magnetic field tracers}
The observables typically used to constrain large scale magnetic fields are the RM structure function and the RM radial profiles. The former is defined as the average mean square difference in RM values between pixels at a given distance dr:
\begin{equation}
    S(dr)=<[RM(r)-RM(r+dr)]^2>,
\end{equation}
and it is mostly used to derive an estimate of the power spectrum of the magnetic field. Indeed, the trend at smaller scales is related, under some assumptions, to the slope of the magnetic field power spectrum \citep[see e.g.][]{ensslin2003A&A...401..835E}.\\
The RM radial profile is instead used to determine the central strength of the magnetic field and its radial trend.\\
\begin{figure}
    \centering
    \includegraphics[width=0.48\linewidth]{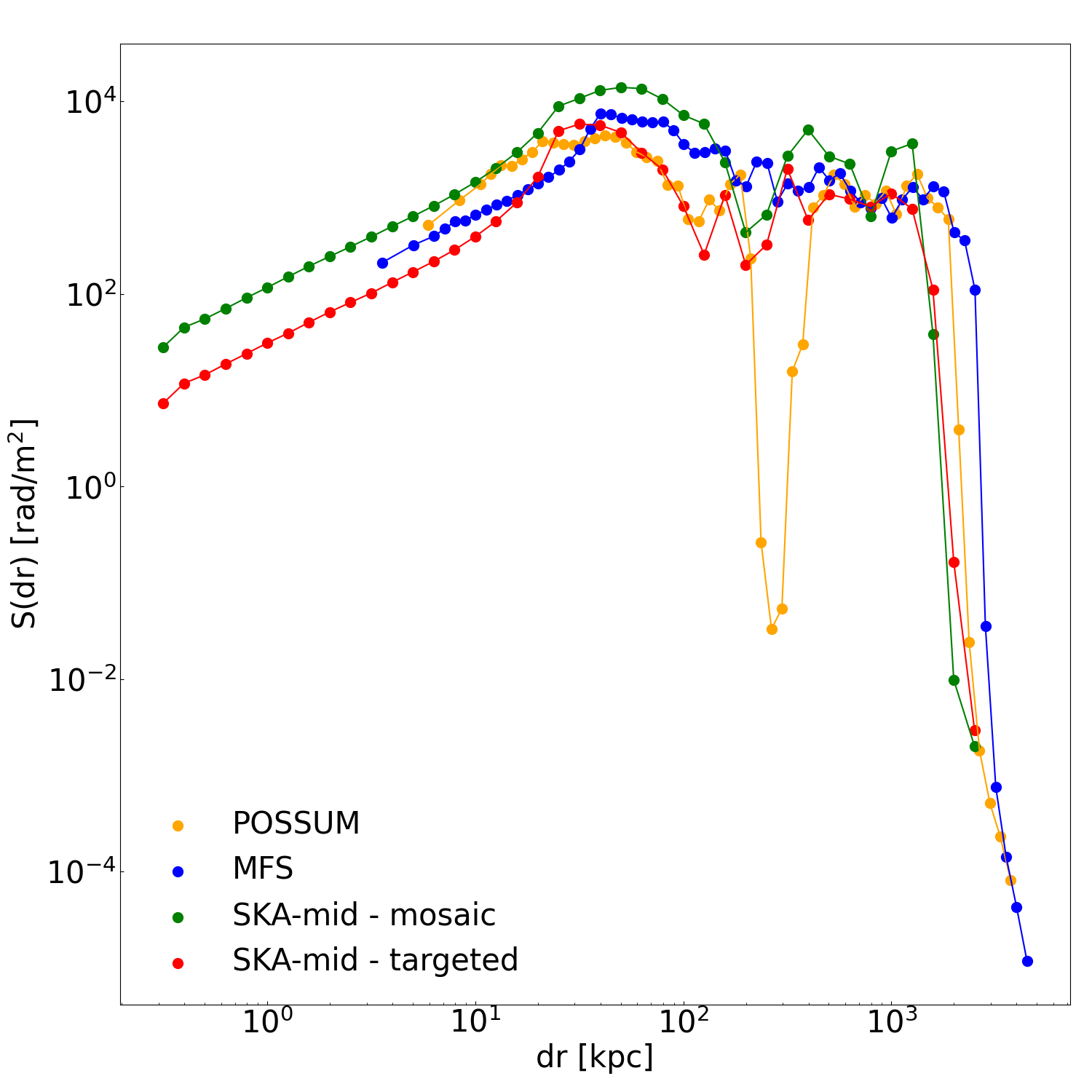}
    \includegraphics[width=0.48\linewidth]{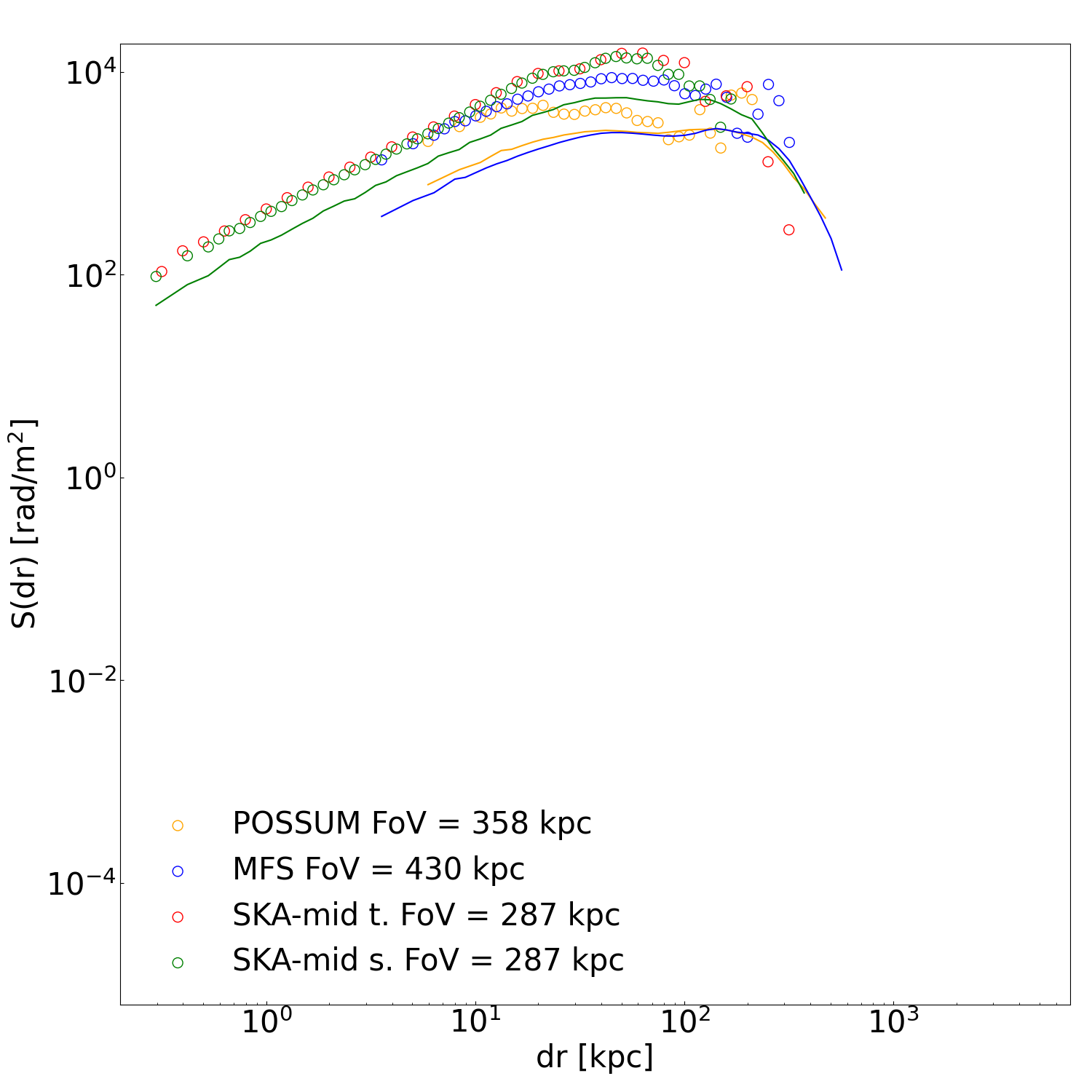}
    \caption{RM structure function calculated from the simulated RM images listed on the bottom left corner (left) and within a smaller FoV reported on the bottom left corner (right). }
    \label{fig:sdr}
\end{figure}

Figure \ref{fig:sdr} presents the RM structure function, computed over separations ranging from 2 pixels up to $\sqrt{2}$ times the FoV that corresponds to 2 Mpc for every image. The left panel compares the different simulated RM images, highlighting the importance of achieving both dense RM grids and high resolution. In particular, the POSSUM and MFS simulations display a significant lack of measurements below distances of 5 and 3\,kpc, respectively. Moreover, at small distances, the POSSUM and MFS data show, respectively, a steeper and a flatter trend with respect to what observed in the SKA-mid simulations. A discrepancy exists between the structure functions of the mosaic and the targeted SKA-mid observations; the latter exhibits lower values at small scales and a continuous trend peaking at lower distances ($\sim$50,kpc). This difference arises because targeted observations provide a higher sampling density compared to mosaic data, ensuring a more robust reconstruction of the magnetic field power spectrum.\\
The right panel of the same figure shows as empty dots the RM structure function calculated over a smaller FoV (reported in the bottom left corner of the Figure) centered on the galaxy cluster core. Notably, the slope of the RM structure function at small separations (<10kpc) remains independent of the chosen FoV in the case of the SKA-mid observations. The presence of an extended polarized source at the cluster center is a key factor to determine the magnetic field properties with high accuracy. Since this is not always the case, it will be important to perform targeted studies to maximize the detection of polarized sources that can probe the cluster RM. While the trend of the structure function does not change in the SKA data, we can see a variation in the case of the MFS and POSSUM observations. Moreover, the separation at which we start seeing a deviation from the continuous trend, that is indicative of the maximum scale of magnetic field fluctuations, shifts towards larger values, in all the cases but the POSSUM survey. 
Solid lines are the RM structure functions computed from the simulated RM images, before the blanking. The green line represents both the SKA-mid mosaic and the targeted observations. Going from the POSSUM to the SKA-mid observation we see a better agreement between the data and the theoretical RM structure function.
The only difference among the SKA-mid targeted, MFS, and POSSUM simulated RM structure functions computed within the smallest FoV is due to the resolution, since within these areas there is only the central extended source. This means that the presence of extended sources and sufficient resolution are the critical factors in accurately reconstructing the RM structure function and in particular, in inferring the slope of the magnetic field power spectrum and the dissipation scale. However, wide FoV are crucial to determine the maximum scale of magnetic field fluctuation.
%
\begin{figure}
    \centering
    \includegraphics[width=0.6\linewidth]{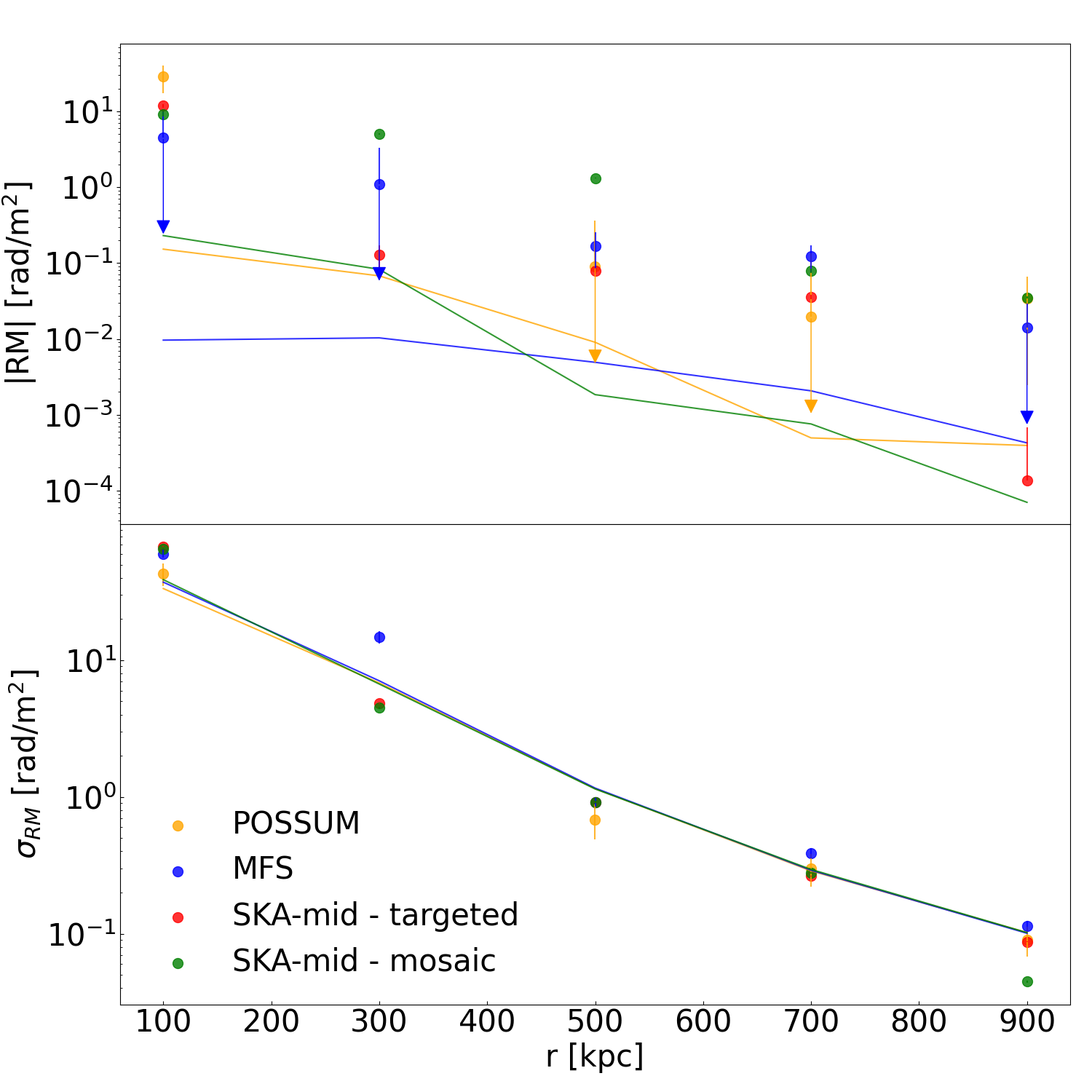}
    \caption{Absolute mean (top) and standard deviation (bottom) RM radial profiles calculated from the simulated RM images listed on the bottom left corner in the top panel. }
    \label{fig:rmp}
\end{figure}

The absolute mean and standard deviation RM profiles are shown in Fig. \ref{fig:rmp}. These quantities were computed in radial bins of 200\,kpc from the cluster center out to a distance of 1 Mpc, excluding bins with fewer than three independent resolution elements. Solid lines refer to the theoretical images that are the simulated images before the blanking, while dots represent the values computed in the simulated images that take into account the sensitivity levels of the observations. The theoretical RM profiles corresponding to the SKA-mid observations are represented in green, since we used the same image for both the mosaic and targeted observation.\\
At the cluster center, the central extended source provides a good sampling of the RM and we see a good match among the four simulations, especially in terms of RM standard deviation. Going towards the external annuli, the area of each radial bin increases and this implies an increasing number of polarized sources as a function of the distance. While the MFS and SKA-mid simulations are not affected by such bias, the POSSUM simulation struggles in sampling the RM at small distances from the cluster center (second radial bin at $\sim$100\,kpc) due to its lower sensitivity.
The differences among the simulations are not statistically significant, with p-values of 0.572 and 0.990 for the absolute mean and standard deviation RM profiles, respectively, computed considering the radial bins sampled by all the simulations. However, the reduced $\chi^2$ values computed as between the data and the theoretical mean RM profiles are 1384, 571, 422, and 160 for the POSSUM, MFS, SKA-mid mosaic, and targeted observations, respectively. This demonstrates that magnetic field measurements based on the mean RM profile require dense RM grids, which can be provided by targeted observations with SKA-mid.

\section{Conclusions}
This study has updated the expected RM grid density values considering the SKA-mid AA4 performance and developed a framework to evaluate the telescope’s potential for studying intracluster and intragroup magnetic fields. We investigated two observational approaches--the SKA-mid mosaic made of 15-minute pointings and a 10--hour targeted observation--and compared these with recent dense RM grids from ASKAP (POSSUM) and MeerKAT (MFS).

A critical finding is the importance of extended sources located near the cluster center to probe the smallest spatial scales, and therefore to constrain the magnetic field power spectrum and its dissipation scale. Our analysis based on the RM structure function shows that high angular resolution, such as that achievable with SKA-mid, is a key factor for accurately determining the magnetic field structure. Resolution proves to be more decisive than sensitivity in this context. Conversely, for capturing the large-scale fluctuations and the radial trends in magnetic field strength, a dense RM grid remains essential as highlighted by the RM structure functions and radial RM profiles. It is worth noting that the RM structure function is dominated by the simulated radio source at the cluster center, a condition that is not always met. This underlines the importance of carrying out deep observations to maximize RM sampling.

Notably, we have neglected the presence of asymmetries in the intracluster medium that are both observed and predicted by numerical simulations as a consequence of hierarchical accretion processes. Therefore, the high sensitivity of SKA-mid will be a game changer in revealing and characterizing such complex features.

We also emphasize that the equation used to estimate the number of polarized sources aligns well with detections obtained by the MFS survey at 13 arcsecond resolution. However, given the improved resolution of SKA-mid, this number might be an underestimate. Importantly, our estimates do not yet incorporate noise fluctuations or internal depolarization effects, which likely reduce the actual detectable polarized sources.

Looking forward, planning targeted studies with SKA-mid will be crucial to fully assess its performance and guide observational strategies. The SKA-mid instrument promises to deliver spatially resolved RM measurements, which will be invaluable for disentangling the complex interactions between magnetic fields and the dynamic processes that govern galaxy clusters and their embedded galaxies.

This work underscores the transformative potential of SKA-mid in advancing our understanding of cosmic magnetism on multiple scales.

\bibliographystyle{abbrvnat-maxbibnames4}
\bibliography{chapter}

\end{document}